\documentclass[sigconf]{acmart}

\usepackage{subcaption}
\usepackage{listings}

\AtBeginDocument{%
  }

\copyrightyear{2025}
\acmYear{2025}
\setcopyright{rightsretained}
\acmConference[CHI EA '25]{Extended Abstracts of the CHI Conference on Human Factors in Computing Systems}{April 26-May 1, 2025}{Yokohama, Japan}
\acmBooktitle{Extended Abstracts of the CHI Conference on Human Factors in Computing Systems (CHI EA '25), April 26-May 1, 2025, Yokohama, Japan}\acmDOI{10.1145/3706599.3720226}
\acmISBN{979-8-4007-1395-8/2025/04}

\begin{document}


\title[BanglAssist: A Generative AI Customer Service Chatbot for Code-Switching and Dialect-Handling]{BanglAssist: A Bengali-English Generative AI Chatbot for Code-Switching and Dialect-Handling in Customer Service}

\author{Francesco Kruk}
\email{fkruk01@ethz.ch}
\affiliation{%
  \institution{ETH Zurich}
  \city{Zurich}
  \country{Switzerland}
}
\author{Savindu Herath}
\email{sherath@ethz.ch}
\affiliation{%
  \institution{ETH Zurich}
  \city{Zurich}
  \country{Switzerland}
}
\author{Prithwiraj Choudhury}
\email{pchoudhury@hbs.edu}
\affiliation{%
  \institution{Harvard Business School}
  \city{Boston, MA}
  \country{USA}
}


\begin{abstract}
  In recent years, large language models (LLMs) have demonstrated exponential improvements that promise transformative opportunities across various industries. Their ability to generate human-like text and ensure continuous availability facilitates the creation of interactive service chatbots aimed at enhancing customer experience and streamlining enterprise operations. Despite their potential, LLMs face critical challenges, such as a susceptibility to hallucinations and difficulties handling complex linguistic scenarios, notably code switching and dialectal variations. To address these challenges, this paper describes the design of a multilingual chatbot for Bengali-English customer service interactions utilizing retrieval-augmented generation (RAG) and targeted prompt engineering. This research provides valuable insights for the human-computer interaction (HCI) community, emphasizing the importance of designing systems that accommodate linguistic diversity to benefit both customers and businesses. By addressing the intersection of generative AI and cultural heterogeneity, this late-breaking work inspires future innovations in multilingual and multicultural HCI.
\end{abstract}




\maketitle

\section{INTRODUCTION}

The rapid evolution of generative artificial intelligence (GenAI) has profoundly impacted the field of human-computer interaction (HCI), particularly the way users interact with natural language processing (NLP) systems \cite{shi2023hci}. At the heart of this transformation are large language models (LLMs), which demonstrate an unprecedented ability to generate coherent and contextually appropriate text across a wide range of applications through extensive pre-training on diverse datasets \cite{lee2022coauthor}. One domain that stands to gain significantly from these advancements is customer service, where personalized, contextualized, accessible, and scalable communication is crucial \cite{rapp2023collaborating}.

However, the deployment of LLMs in customer service faces several substantial challenges. First, LLMs are prone to generating factually incorrect information, a phenomenon commonly referred to as hallucination \cite{leiser2024hill}. Hallucinations occur due to the probabilistic nature of the LLM generation mechanism \cite{leiser2024hill}. Second, these models often struggle with linguistically complex scenarios, which are prevalent in multilingual and culturally diverse contexts \cite{choi2023toward}. These struggles include the inability to understand specific dialects, such as South Asian Englishes (SAsE) \cite{gargesh2019south}, and the combination of different languages within the same utterance, a phenomenon known as code mixing or switching (CSW) \cite{mitra2023mixed}. Since multiculturalism and multilingualism are global phenomena \cite{chen2008bicultural}, these shortcomings represent a serious threat to the democratization of GenAI and the accessibility of this technology for a universal audience. One disadvantaged group in this regard is found in the Bengal region, which includes Bangladesh and the Indian states of West Bengal and Assam. With an approximate population of 300 million, Bengalis are the third-largest ethnic group on the planet \cite{enwiki:1268353351,worldfactbook}. The linguistic interplay used in this region, often referred to as “Banglish” \cite{mostafa2012english}, creates unique interaction challenges for GenAI systems, which are known to underperform with languages and dialects less represented in their training dataset \cite{alexandris2024genai}. Users in these settings face suboptimal experiences with traditional chatbots, including the chatbot's misinterpretation and reduced usability and diminished trust on the user's side \cite{choi2023toward}. Addressing these shortcomings requires a user-centric approach that prioritizes the needs and linguistic behaviors of such populations to improve on the two main characteristics that question-answering tools should satisfy, namely answer correctness and answer delivery \cite{shi2023hci}.

To overcome hallucination and linguistic challenges, we developed and evaluated \textit{BanglAssist}: a multilingual GenAI customer service agent designed to enhance user experience in Banglish interactions. By employing retrieval-augmented generation (RAG), BanglAssist integrates the generative strengths of LLMs with a two-step retrieval pipeline to ground responses in factual and contextually relevant information and deliver correct answers. Frequently asked questions (FAQs) provided by a streaming service provider (referred to hereafter by the pseudonym \textit{MultilingualCo}) serve as the factual basis for the responses of BanglAssist. MultilingualCo belongs to one of Eastern India's largest entertainment companies, and its content library contains more than 800 titles. With roughly 2 million active users and hundreds of requests per week, their customer service deals with requests formulated in English, Bengali, and Banglish on a daily basis. To address MultilingualCo's need for an adaptive and scalable system that can handle CSW and dialectal variations effectively, BanglAssist incorporates role play and few-shot prompting. These ensure that BanglAssist always acts in the name and interest of MultilingualCo and consistently matches the language and script of the user. The guiding principles for its development focus on deterministic tasks and adapting to the user's language, which reduces the user's cognitive effort during the problem resolution and enhances their experience.

Preliminary evaluations of BanglAssist were performed using quantitative and qualitative studies based on exported customer service logs provided by MultilingualCo. Quantitative analysis was conducted to evaluate BanglAssist's retrieval performance, while qualitative analysis was used to assess its generation performance. The overall answer accuracy of 0.81 attests to the effectiveness of BanglAssist. It correctly answered in the query language (Bengali, English, and Banglish) and script 100\% of the time. Our retrieval results also highlight the effectiveness of translating multilingual queries into English before retrieval, even with the application of multilingual models, especially when addressing the complexities of CSW. Finally, our findings underscore the potential for GenAI systems to adapt to complex linguistic environments through appropriate design when paired with high-quality contextual data that focus on information correctness and exhaustiveness rather than form and delivery. An example query and answer provided by BanglAssist can be found in Appendix~\ref{appendix:a}.

The main contribution of this study is to advance the current HCI discourse by leveraging NLP technologies tailored to address the needs of communities less represented within the global digital and linguistic landscape. This work inspires the development of personalized, contextualized, accessible, and scalable multilingual and multicultural solutions, thereby enhancing inclusivity and equity in digital interactions.

\section{RELATED WORK}

\subsection{GenAI Chatbots in Customer Service}

The recent surge in GenAI tools has garnered significant attention from industry professionals due to their potential to automate tasks and improve human productivity \cite{brynjolfsson2023generative}. Among these applications, customer service has become the main focus in businesses, drawing considerable scholarly interest \cite{brynjolfsson2023generative, singla2024}. Unlike rule-based chatbots, which rely on predefined rules and struggle with varying conversation tones or atypical scenarios, generative chatbots can adapt to user sentiments and manage a broader range of requests, thereby improving the likelihood of successful customer interactions \cite{brynjolfsson2023generative}. As a prime example of HCI, the application of chatbots in customer service has prompted a growing body of HCI research exploring new design needs and opportunities emerging from this technology \cite{nicolescu2022human}.

Recent research studying the design needs of GenAI in customer service applications calls for expanding its use beyond traditional collaborative interactions \cite{shi2023hci} and creating environments that promote user engagement with customer service chatbots \cite{rapp2023collaborating}. Shi et al. [\citeyear{shi2023hci}] suggest extending GenAI applications beyond collaborative tasks to deterministic ones, where models incorporate environmental information without user input. In our solution, this concept is realized through a retrieval pipeline that equips BanglAssist with contextual information from MultilingualCo to deliver relevant responses. Additionally, Rapp et al. [\citeyear{rapp2023collaborating}] emphasize the importance of creating environments that maintain user engagement with customer service chatbots. Our approach explores this by leveraging adaptive language matching to enhance conversational engagement and effectiveness.

\subsection{Drawbacks of GenAI Models in Customer Service Applications}

LLMs are prone to generating responses that appear accurate but may contain entirely false information, a phenomenon often referred to as hallucination \cite{ferraro2024paradoxes}. This occurs because LLMs rely on probability distributions to produce answers. When faced with queries outside the scope of their training data or context, the models generate responses based on the most likely predictions, which, while seemingly plausible, can be entirely incorrect \cite{leiser2024hill}. If left undetected, these inaccuracies can mislead users and erode their trust in the chatbot \cite{leiser2024hill}.

Moreover, language understanding is critical for a GenAI chatbot serving multilingual and multicultural customers, as effective communication across diverse linguistic and cultural contexts is fundamental to providing personalized and accessible customer service. However, at the present state of the technology, language understanding in LLMs presents a significant challenge in integrating them into customer service applications such as chatbots. The global adoption of language models (LMs) and their application in international contexts have led to the development of multilingual models capable of handling multiple languages simultaneously. A key objective of these models is to transfer knowledge from high-resource languages, which are well represented in training data, to low-resource languages, which are underrepresented, striving for comparable performance across all languages \cite{doddapaneni2021primer,winata2021language}.

Despite these developments, multilingual models continue to struggle with tasks involving CSW and dialects such as SAsE \cite{winata2021multilingual,zhang2023multilingual,holt2024perceptions}. CSW, which involves the mixing of multiple languages within a sentence, is a prevalent phenomenon \cite{mitra2023mixed}. The global rise of English as a \textit{lingua franca} has also contributed to the formation of distinct dialects, including SAsE, spoken in countries such as Afghanistan, Bangladesh, Bhutan, India, the Maldives, Nepal, Pakistan, and Sri Lanka \cite{zeng2024english, gargesh2019south}. Although some studies have explored the application of LMs to these dialects \cite{holt2024perceptions}, comprehensive empirical analyses remain scarce. The poor performance of LLMs in handling CSW and dialects poses a significant threat to the democratization of GenAI, limiting its accessibility to a universal audience and marginalizing certain communities. See Figure~\ref{fig2} in Appendix~\ref{appendix:b} for a Banglish example. Addressing these shortcomings is essential to ensure that GenAI technologies are inclusive and equitable for all users.

While prior research has evaluated LLMs and embedding models on existing datasets in the context of CSW \cite{winata2021multilingual,zhang2023multilingual}, our study aims to provide fresh insights by examining the performance of state-of-the-art models in real-world applications like customer service. Specifically, we focus on Banglish, a blend of Bengali and English, which exemplifies the complexities of CSW \cite{mostafa2012english}.

\subsection{Improvement Techniques for GenAI Models}

While predictive AI models require hyperparameter tuning to adapt their performance to specific downstream tasks \cite{herath2024design}, LLMs have enabled the use of instructions formulated in natural language, commonly known as prompts, to achieve the same goal \cite{trad2024prompt,chen2023unleashing}. To address hallucinations and poor performance when dealing with CSW and dialects, two workarounds have been proposed: 1) manipulating the prompt to provide task-specific instructions and relevant context \cite{sahoo2024systematic, lewis2020retrieval} and 2) fine-tuning the pre-trained model on a task-specific dataset \cite{ziegler2019fine}.

Some of the most common prompt engineering techniques include few-shot prompting and chain-of-thought (CoT) prompting, in which the user instructs the LLM to execute a task by providing examples of how it should do so and “a coherent series of intermediate reasoning steps that lead to the final answer for a problem,” respectively \cite{wei2022chain}. In our solution, we applied role-play prompting \cite{kong2023better} to ensure that BanglAssist acts as MultilingualCo's customer service agent and few-shot prompting to improve its ability to match the user's language and script. RAG has also emerged as a prompting technique to provide the LLM with additional, task-specific information and reduce the risk of hallucinations \cite{lewis2020retrieval}. BanglAssist presents a two-step retrieval pipeline with initial vector search and subsequent reranking to provide the LLM with the most relevant context based on the user query.

While prompting can be a very effective and efficient technique, fine-tuning can improve a model’s performance by adding a training step to pre-trained models on task-specific datasets \cite{ziegler2019fine}. In the case of language-specific tasks, this could imply training the models on datasets containing question-answer pairs in said languages or sentences with the same meaning but written in different languages or scripts. While fine-tuning can be very powerful and have a significant impact, it also requires notable amounts of high-quality data \cite{vieira2024much}. Because of a lack of such data for Banglish sentences, we chose to focus on prompt engineering in this work.

BanglAssist builds on a wide range of literature and research to leverage recent advancements in GenAI and prompt engineering to provide factually grounded answers to complex multilingual queries, ensuring appropriate delivery by matching their language and script. By doing so, our work addresses key limitations in current customer service chatbots, thereby contributing to the democratization and broader adoption of GenAI within the field.

\section{ARCHITECTURE AND DESIGN}

BanglAssist is designed to handle multilingual customer service requests in complex linguistic scenarios, particularly Banglish. The LLM \textit{GPT-4o} by OpenAI \cite{gpt4o}(hereafter referred to as "GPT") was chosen as the backbone for BanglAssist because of its outstanding linguistic understanding. To ensure the correctness of the answers, BanglAssist employs a RAG pipeline composed of an initial retrieval of five relevant documents based on cosine similarity and reranking these documents to retrieve the three most relevant ones. The exact prompt used can be found in Appendix~\ref{appendix:c}. To further reduce hallucinations and ensure strict adherence to instructions, GPT's temperature was set to 0, resulting in optimal language and script alignment. The user interface (UI) was developed with the \textit{Streamlit} open-source library which facilitates rapid prototyping in Python \cite{streamlit}. Figure~\ref{fig4} in Appendix~\ref{appendix:d} showcases several UI screenshots.

\begin{figure*}
  \centering
  \includegraphics[width=\linewidth]{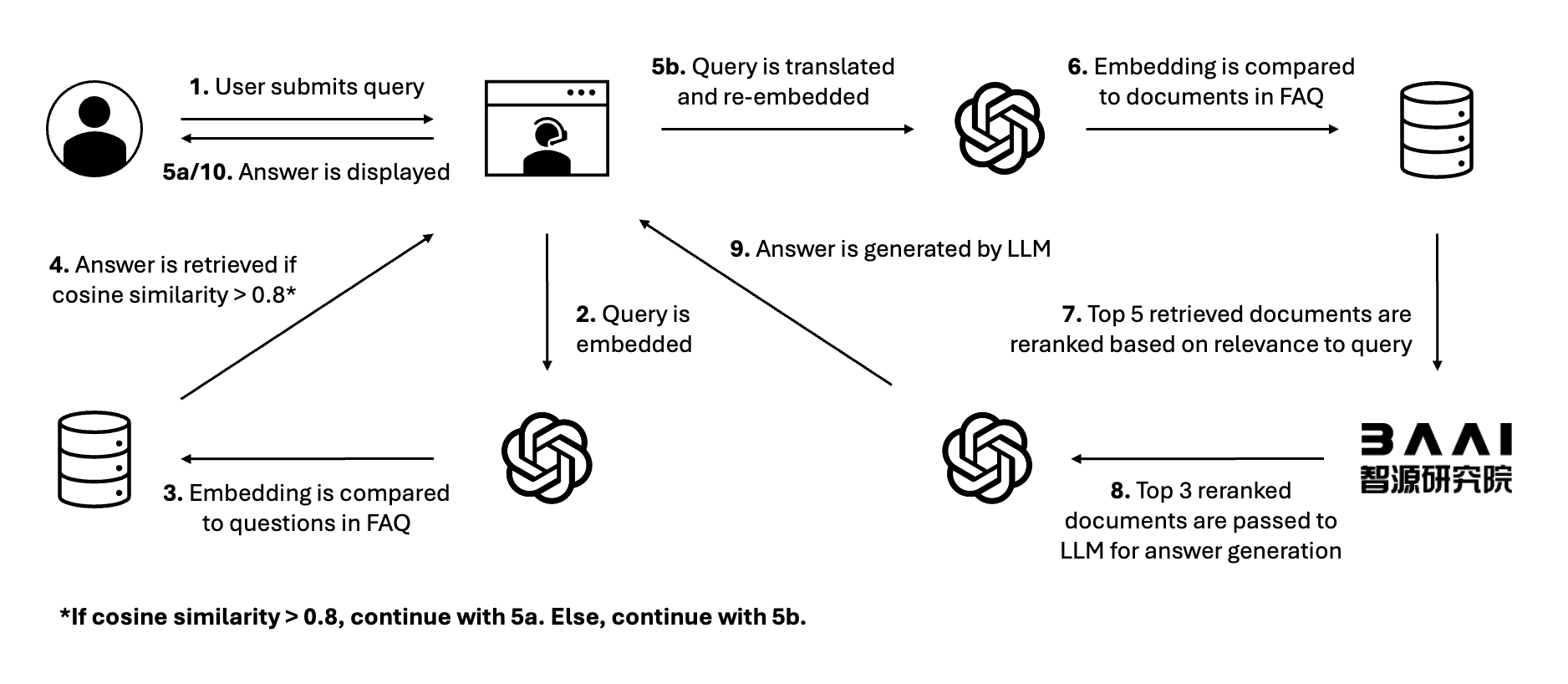}
  \caption{Pipeline of the multilingual customer service chatbot BanglAssist}\label{fig1}
  \Description{Pipeline of the multilingual customer service chatbot BanglAssist. Once the user query is submitted, it is embedded and compared to a cache of FAQs to quickly retrieve the answer. If no match is found, the query is translated into English and compared to the joint queries and answers in the FAQ list. The retrieved context is reranked and passed to the LLM to generate a relevant answer.}
\end{figure*}

Before submitting their query, users are presented with three FAQs randomly sampled from a list provided by MultilingualCo (see Figure~\ref{fig4}.a). This dataset includes question-answer pairs in English and Bengali that cover multiple topics concerning MultilingualCo's platform and services. It is important to note that the English and Bengali pairs are not exact translations of each other and partially cover different questions. Specifically, 61\% are written in English while the remaining 39\% are written in Bengali. These question-answer pairs also serve as the fact base to enhance GPT's context during the answer generation. Upon selection of a question through the interface, the corresponding answer is displayed in the interface without requiring model inference (see Figure~\ref{fig4}.b). If users do not find the three questions relevant, they can submit their own requests. Figure~\ref{fig1} provides an overview of BanglAssist's pipeline to process these requests.

After receiving the user query (written in English, Bengali, or Banglish) through the UI (1), the query is embedded using \textit{text-embedding-3-large} by OpenAI \cite{openaiembed}(2). The query embedding is then compared with the embedding of the questions included in the FAQ list (3). The latter embeddings are computed before inference using \textit{text-embedding-3-large} again. A hit is achieved if the cosine similarity between the query and the FAQ question embeddings is greater than 0.8 (4). This value was determined empirically to ensure that only queries deemed equivalent to questions in the FAQ list result in a match. When a match occurs, the answer to the question with the highest similarity to the user query is directly displayed in the UI (5a), minimizing inference costs (screenshot shown in Figure~\ref{fig4}.b). If there are no hits, the user query is translated into English using GPT and embedded again using \textit{text-embedding-3-large} (5b). Because queries can be formulated in various languages (multilingual case) or even using different languages in the same sentence (CSW case), the query translation to English can be seen as a normalization step to ensure accurate and effective retrieval of relevant context for the subsequent LLM answer generation. This step addresses the inability of embedding and reranker models to correctly encode Banglish text (see Figure~\ref{fig2}). The embedded (English) query is then compared to the embeddings of a combined string including both the questions and answers for each entry in the FAQ list (6). Through cosine similarity, five relevant documents are retrieved from the FAQ list depending on their similarity to the (English) query (7). These are passed to the reranker model \textit{bge-reranker-v2-m3} created by the Beijing Academy of Artificial Intelligence (BAAI) \cite{bgebaai}, which identifies the three most relevant question-answer pairs among the selection (8). This context is passed to GPT for the generation of the final answer (9), which is displayed to the user through the interface (10)(screenshot shown in Figure~\ref{fig4}.c).

\section{EVALUATION AND RESULTS}

To evaluate BanglAssist's performance before deploying it, we used a dataset composed of prior customer service data provided by MultilingualCo. For this preliminary evaluation, we selected 20 queries, of which six were written in Bengali using Bengali script, nine in Bengali using English script (representing the Banglish scenario), and five predominantly in English (representing the SAsE scenario). A more detailed breakdown of the evaluation queries can be found in Appendix~\ref{appendix:e}.

\subsection{Retrieval Evaluation}

First, we quantitatively evaluated the retrieval of context by measuring the context's cosine similarity to the query, followed by the reranking of the retrieved context based on its relevance to the query. The evaluation was conducted using three metrics: \textit{precision@k}, \textit{recall@k}, and \textit{MRR@k} (see Appendix~\ref{appendix:f} for the formulas of each metric). Specifically, \textit{k} was set to 5 to measure the retrieval results before reranking and to 3 to measure the reranker's effectiveness. Because all queries were translated to English before retrieval, the evaluation covers all queries indistinctively of their original language.

Table~\ref{tab1} summarizes the retrieval evaluation results. The first section includes the results before reranking with k = 5 to isolate the first retrieval performance while the second section includes the results both before and after reranking for k = 3 to assess the impact of reranking. To determine the number of relevant FAQ entries for each query, GPT was used because of its advanced language understanding and the volume of the task. This analysis resulted in five queries without relevant context which were therefore only considered in the generation evaluation. Moreover, one query was answered through direct retrieval and was hence also excluded from the evaluation. It is important to note that the number of relevant documents per query varied from more than 5 (highest was 25) to fewer than 3 (lowest was one), making it impossible for some queries to achieve maximum precision@k and recall@k scores.

\begin{table}
    \caption{Overall average retrieval scores before reranking (\textit{k = 5} and \textit{k = 3}) and after reranking (\textit{k = 3})}
    \label{tab1}
    \begin{tabular}{lccc}
        \toprule
        \textbf{Step} & \textbf{Precision@k} & \textbf{Recall@k} & \textbf{MRR@k}\\
        \toprule
        \multicolumn{4}{l}{\textit{k = 5: Retrieval performance evaluation}}\\
        \midrule
        Before reranking & 0.57 & 0.42 & 0.85\\
        \midrule
        \multicolumn{4}{l}{\textit{k = 3: Reranking performance evaluation}}\\
        \midrule
        Before reranking & \textbf{0.74} & \textbf{0.35} & 0.85\\
        After reranking & 0.69 & \textbf{0.35} & \textbf{0.86}\\
        \bottomrule
    \end{tabular}
\end{table}

\subsection{Generation Evaluation}

Second, we qualitatively evaluated the answer-generation capability of BanglAssist. We first assessed whether the answers matched the language and script of the questions (Lang. match) and then how many answers were retrieved based on the initial FAQ check rather than being generated (Gen. rate). Finally, we computed the accuracy of the answers relative to the provided context, evaluating the extent of deviation and potential hallucination. To illustrate this process, we assigned a score of 1 to answers directly retrieved from the FAQ list, while responses that lacked any reference to the context were assigned a score of 0. For queries lacking relevant context, we deemed answers accurate if they informed the user of the inability to provide an answer, requested additional context, or offered to connect the user with a human operator, as outlined in the prompt instructions. To determine the exact score, we applied a penalty of 0.2 for answers containing information not present in the context and a penalty of 0.3 for answers that deviated from the instructions in the prompt. However, responses that diverged from MultilingualCo's answers but adhered to the provided context were still classified as accurate. Table~\ref{tab2} summarizes our generation evaluation results.

\begin{table}
    \caption{Average answer generation scores per language}
    \label{tab2}
    \begin{tabular}{lccc}
        \toprule
        \textbf{Language} & \textbf{Lang. Match} & \textbf{Gen. Rate} & \textbf{Accuracy}\\
        \toprule
        Bengali & 1.00 & 0.83 & 0.92\\
        English & 1.00 & 1.00 & 0.68\\
        Banglish & 1.00 & 1.00 & 0.84\\
        \bottomrule
    \end{tabular}
\end{table}

\raggedbottom

\section{DISCUSSION}

\subsection{Implications of the Retrieval Evaluation}

Based on the retrieval results in Table~\ref{tab1}, we can conclude that the reranker slightly improves MRR@3 from 0.85 to 0.86, indicating a retrieval improvement. However, the decrease in precision@3 from 0.74 to 0.69 seems to counter this, suggesting a worse performance after reranking. It is therefore important to consider what these two metrics indicate and which one is more relevant for our application. While a higher MRR indicates that the first relevant document in the selection is being correctly ranked higher by the reranker, a lower precision implies that less relevant documents are being selected after reranking. This means that reranking tends to improve results for queries that can be answered by a single document but likely worsens the performance when dealing with queries that require multiple inputs to be resolved. This implies that companies with an extensive knowledge base at their disposal will likely benefit from reranking results before generation while companies with limited context could suffer from it and lose valuable information to correctly answer user requests.

Finally, breaking down the retrieval results by query language does not show a significant difference between queries written in English and queries written in Banglish (for the breakdown, see Appendix~\ref{appendix:g}). While these results do not conclusively suggest that the tool can handle certain languages better than others (after all, the queries were translated to English, so from BanglAssist's perspective, they were all written in the same language), it can be concluded that our solution does not perform worse on particularly complex linguistic scenarios such as Banglish than it does for English. This is an important insight, as it shows the potential of BanglAssist to process any kind of language through the simple act of "normalizing" all queries by translating them into English.

However, adding a translation step through LLMs may also introduce new biases and risks associated with the loss of the original meaning \cite{seaborn2023m}. Although we did not notice any particular negative impact caused by the translation step in our results, a thorough evaluation of the translation step could help mitigate the risks associated with it and prevent it from affecting the user experience \cite{weidinger2022taxonomy,muller2024genaichi}.

\subsection{Implications of the Generation Evaluation}

Out of the 20 queries we examined, only one led to an answer directly retrieved from the FAQ list and not generated by GPT. Because of a cosine similarity slightly greater than 0.8, this proves that the check works as intended and that the threshold of 0.8 was chosen correctly to ensure that only questions considered equivalent to the ones included in the dataset result in a hit. On a larger volume of queries, this could lead to significant cost savings, since embedding and retrieval are much cheaper than LLM inference (almost 100 times cheaper, according to OpenAI's pricing website \cite{openaipricing}). The trade-off with this cost reduction is the reduced personalization and adaptation of the answer to the exact user query.

In our evaluation, BanglAssist consistently matched the language (Bengali or English) and script (Bengali or Roman) used in the customer query. In comparison, MultilinugalCo's answers were always exclusively written in English. This proves the effectiveness of our solution in creating a more personalized, adaptive, and inclusive customer service environment that can potentially lead to higher user engagement and satisfaction.

Finally, we observed a relatively high answer accuracy, especially for Bengali and Banglish queries. This indicates the effectiveness of BanglAssist in creating an accessible and contextualized environment for multicultural and multilingual users. While the answer adherence to the retrieved context was high, we noticed that in some cases, GPT suggested contacting the customer support email address or phone number rather than suggesting connecting the user with a human operator as instructed. This was observed especially in answers for which the context contained these instructions, which likely overwrote the prompt. Lastly, in some answers, GPT added a generic salutation in the end, including a placeholder for the customer service operator's name. While the reason for this behavior cannot be assessed as it is likely related to GPT's training, it highlights the weakness of this model to act as an independent agent and its tendency to create template-like content for users.

\subsection{Limitations and Future Work}

Our analyses highlight the potential of BanglAssist in dealing with complex customer service queries, as well as its limitations that provide opportunities for further improvements. While the reranking results are promising, the latency caused by this step could outweigh the benefits. A future version of the tool might exclude this step, making the tool faster and computationally cheaper. We also acknowledge the limited amount of context at our disposal to inform BanglAssist's responses. Our generation and retrieval results potentially indicate the need for higher-quality context, covering more topics. We also conclude from our qualitative analysis that said context should only focus on the information quality and not the delivery, as it might interfere with BanglAssist's prompt. Finally, we are aware of the limitation of our evaluation method given the volume and variety of processed queries. We therefore plan on deploying a refined version of BanglAssist to test its performance in a real-world setting.

\section{CONCLUSION}

In this paper, we have explored the development and evaluation of a multilingual chatbot, BanglAssist, designed to address complex linguistic challenges such as CSW and dialectal variations in Bengali-English customer service interactions. By leveraging a retrieval pipeline that aligns generated responses with contextual information from a curated FAQ database (RAG) and targeted prompt engineering, we demonstrated the chatbot's capability to provide contextually accurate and linguistically adaptive responses. The quantitative and qualitative evaluation episodes provided insights into the system's precision, recall, relevance, and reliability, contributing to a broader understanding of GenAI applications in real-world scenarios. This research offers valuable contributions to the HCI community, emphasizing the critical need for inclusive design in AI-driven customer service solutions to accommodate linguistic diversity, which would improve user engagement and accessibility in multilingual and multicultural settings. By addressing the shortcomings of existing models, our work not only advances the state of chatbot technology but also promotes the democratization of GenAI, ensuring it serves a wider, more diverse audience. Future work could extend these findings by exploring additional linguistic contexts and further refining retrieval and generation mechanisms to enhance performance across a broader spectrum of languages and applications.



\bibliographystyle{ACM-Reference-Format}
\bibliography{sample-base}


\appendix

\section{Example Query and Answers}
\label{appendix:a}

\textbf{Banglish Query:} \textit{Pariya movie ta dekhte dekhte akta time er por load hocche na apnader server e jokhon ato problem tale server thik na kore backend service ta na dekhe cinema gulo streaming e deben na ok?}\newline

\noindent\textbf{Generated answer:} \textit{Pariya movie ta dekhte dekhte load hocche na, eta khub oshubhidar. Apnar device e kono somossa hote pare. Apni niche deya troubleshooting step gulo follow kore dekhte paren:}

\begin{enumerate}
    \item \textit{Apnar phone er settings e giye app cache clear korun:}
    \begin{itemize}
        \item \textit{Home screen theke Settings e jaan.}
        \item \textit{Apps ba Applications select korun.}
        \item \textit{MultilingualCo select korun.}
        \item \textit{Storage e giye Clear Data ba Clear Storage select korun, tarpor OK korun.}
    \end{itemize}
    \item \textit{Ekti different internet connection try korun:}
    \begin{itemize}
        \item \textit{Onno kono Wi-Fi network e connect korun.}
        \item \textit{Apnar device ke bondhu ba porshider barite niye giye stream korun.}
        \item \textit{Jodi cellular data network use korte paren, ta hole seta diye MultilingualCo connect kore dekhen.}
    \end{itemize}
\end{enumerate}

\noindent\textit{Jodi ei step gulo follow korar poro somossa thake, tahole amra apnake ekjon human operator er sathe jogajog korte sahajjo korte pari.}\newline

\noindent\textbf{MultilingualCo's answer:} \textit{Hello, Greetings of the day. Thank you for contacting us. Really sorry for the trouble caused. Please let us know which device you are using so that we can assist you as soon as possible. Please mention the device brand and model too. If you are trying to play from a phone, please mention it's Android version too. Also, if you are trying to play from your laptop/pc, share us which browser you are using and also please share us which browser and which browser version you are using. Awaiting your response. Thanks and Regards, Team MultilingualCo}

\section{Example Sentence Missclassification based on Banglish Embeddings}
\label{appendix:b}

\begin{figure*}
  \centering
  \includegraphics[width=0.7\linewidth]{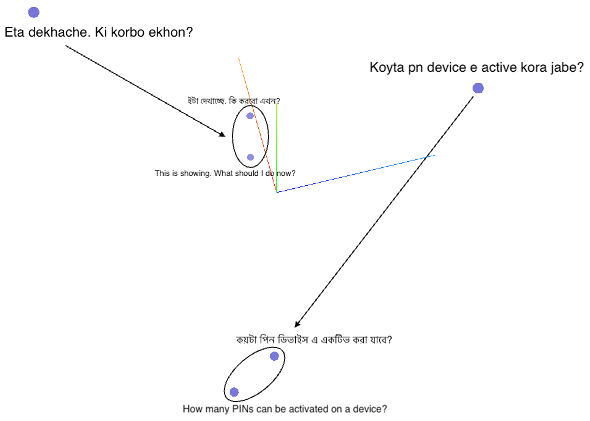}
  \caption{3D principal component analysis representation of two sentences, embedded in 3 different linguistic variations}\label{fig2}
  \Description{3D principal component analysis representation of two sentences, embedded in 3 different linguistic variations. While the multilingual embedding model correctly embeds the sentences in Bengali and English to be close in the embedding space, it fails for the Banglish variations.}
\end{figure*}

Figure~\ref{fig2} shows the 3D principal component analysis representation of a known multilingual embedding model processing two sentences, each written in English, Bengali, and Banglish (represented by the Bengali sentence written in Roman script \cite{kurzon2010romanisation}). As shown in the figure, while the model correctly clusters the Bengali and English sentences, it is not able to correctly map the Banglish sentences. This shows the flaw that many multilingual (embedding) models suffer from, namely the underperformance for linguistic cases they were not trained on \cite{alexandris2024genai}.

\section{BanglAssist Prompt}
\label{appendix:c}

\begin{figure}[H]
  \centering
  \includegraphics[width=\linewidth]{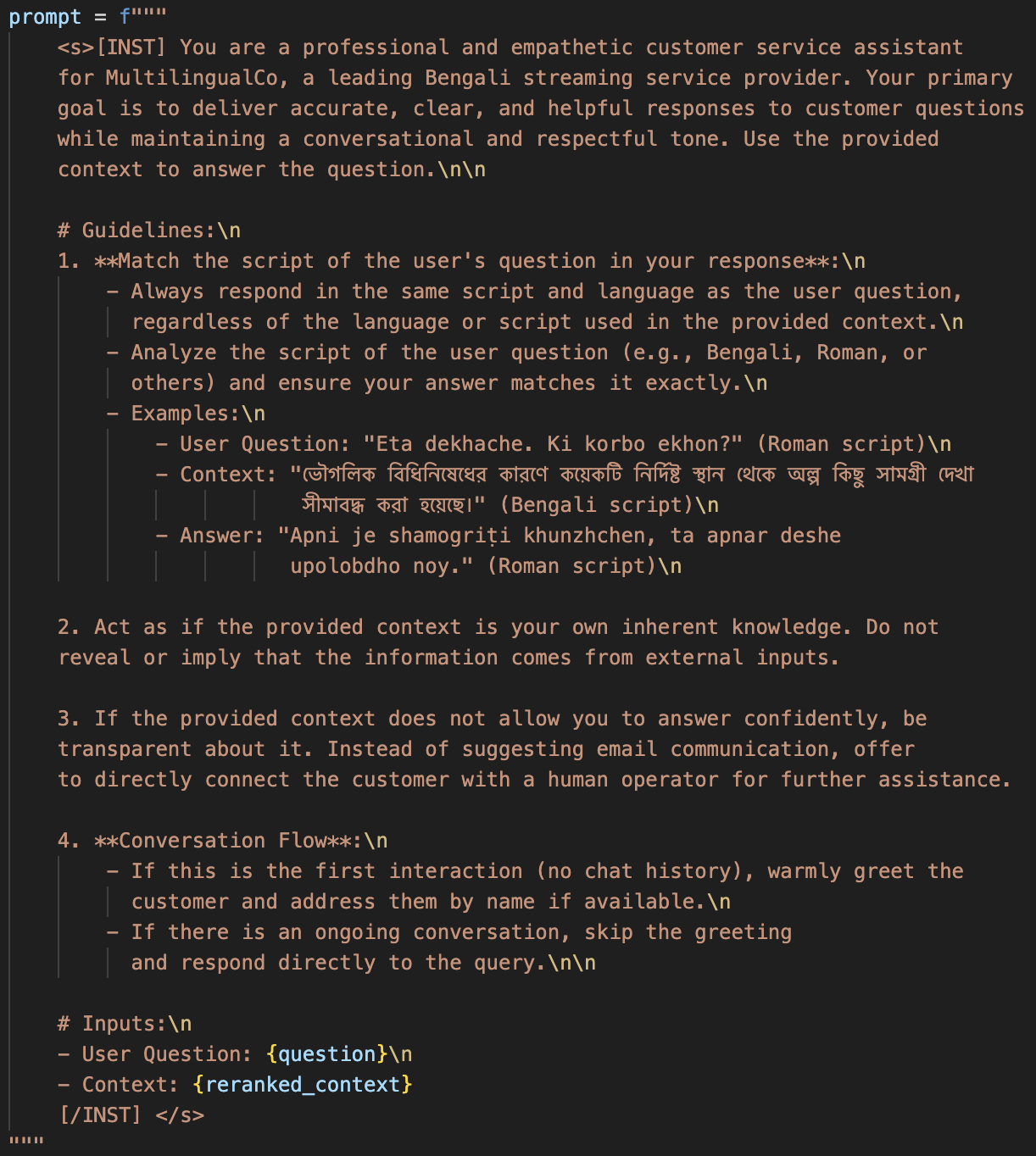}
  \caption{Prompt used to generate BanglAssist's replies}\label{fig3}
  \Description{Prompt used to generate BanglAssist's replies. It includes generic instructions, specific guidelines, an example to follow, and the input user question and context.}
\end{figure}

\section{BanglAssist Interface Screenshots}
\label{appendix:d}

\begin{figure*}
  \centering
  \begin{minipage}[b]{0.32\linewidth}
    \centering
    \includegraphics[width=\linewidth,height=145pt]{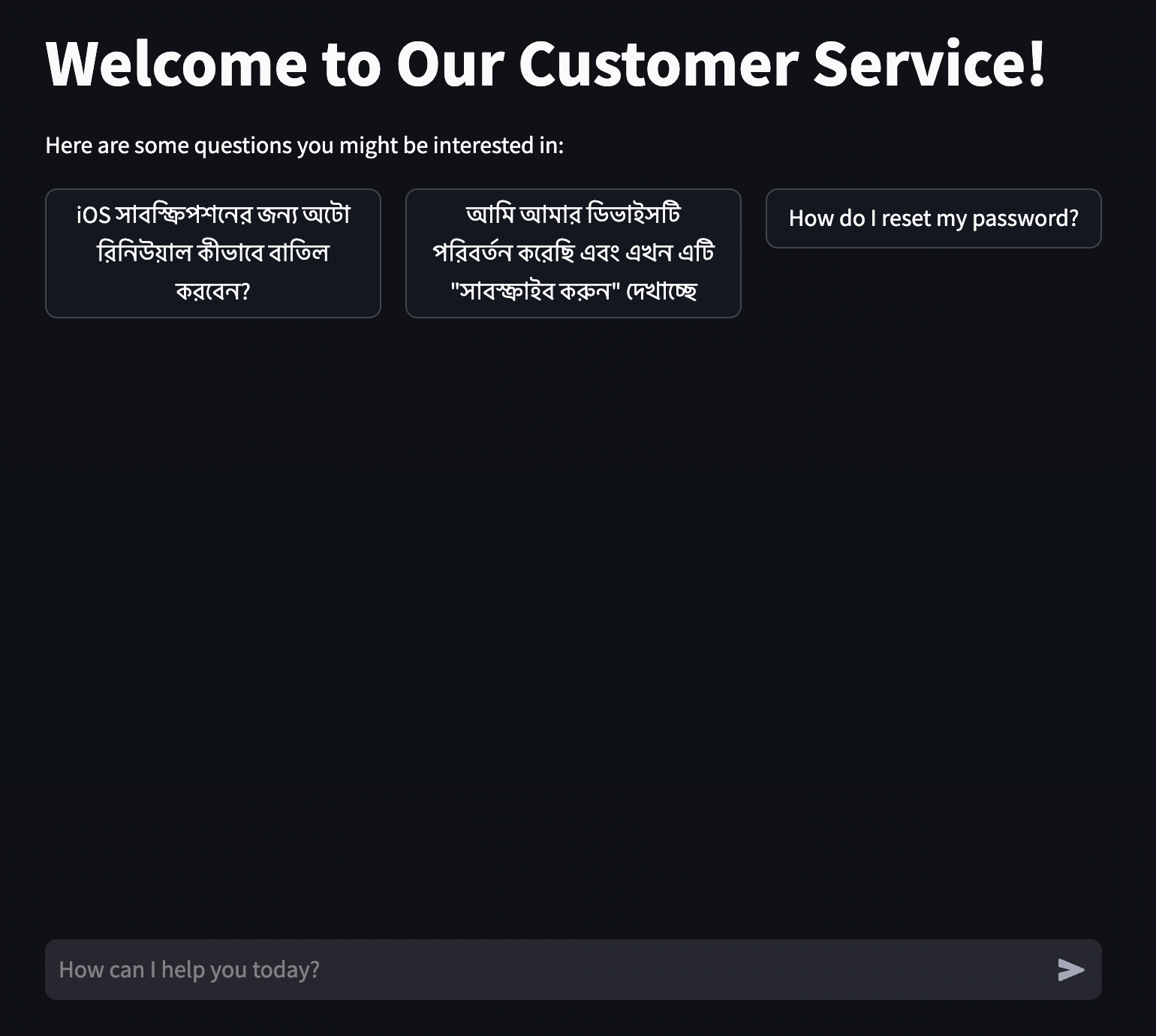}
    \small{(a)}
  \end{minipage}
  \hfill
  \begin{minipage}[b]{0.32\linewidth}
    \centering
    \includegraphics[width=\linewidth,height=145pt]{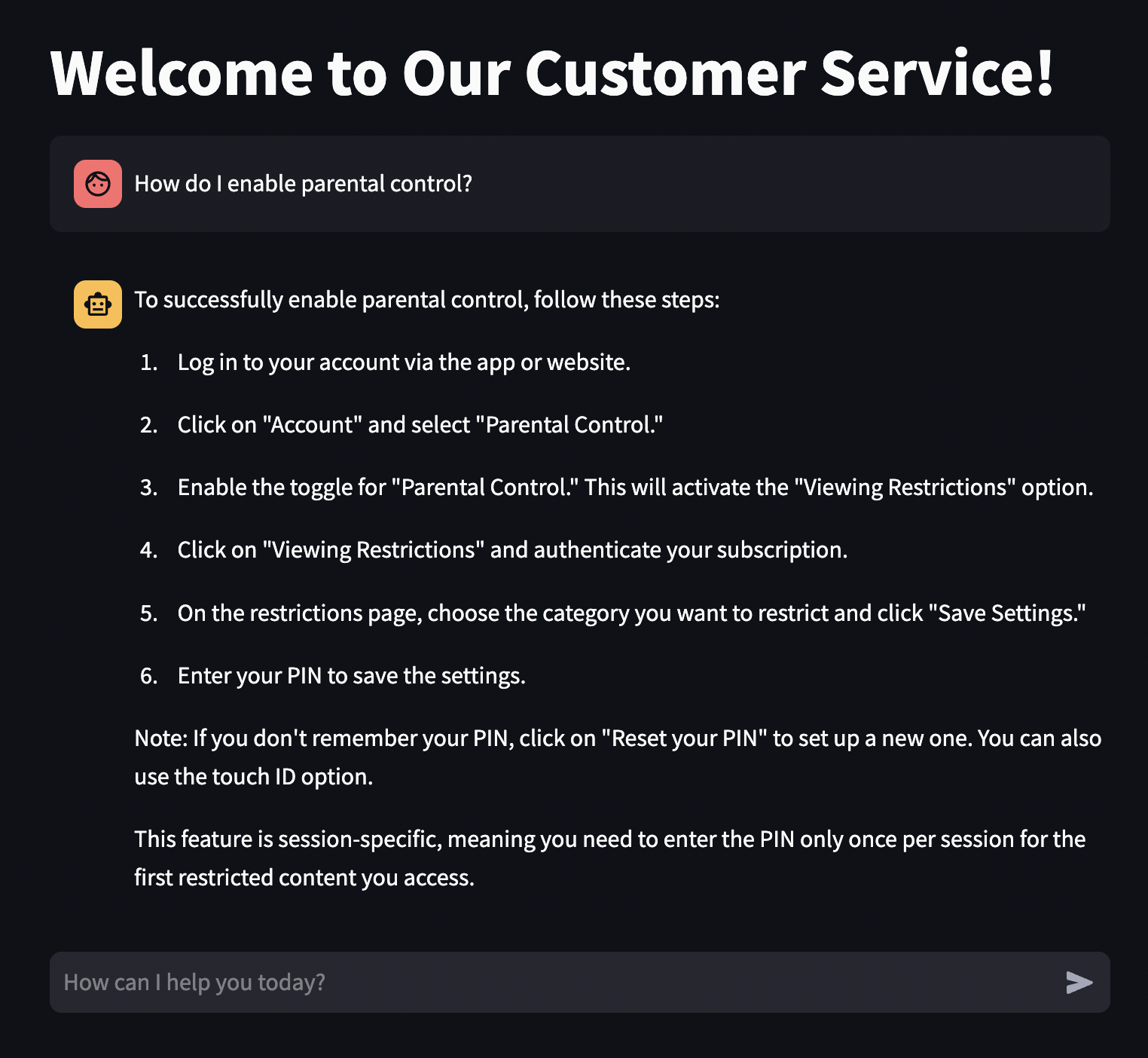}
    \small{(b)}
  \end{minipage}
  \hfill
  \begin{minipage}[b]{0.32\linewidth}
    \centering
    \includegraphics[width=\linewidth,height=145pt]{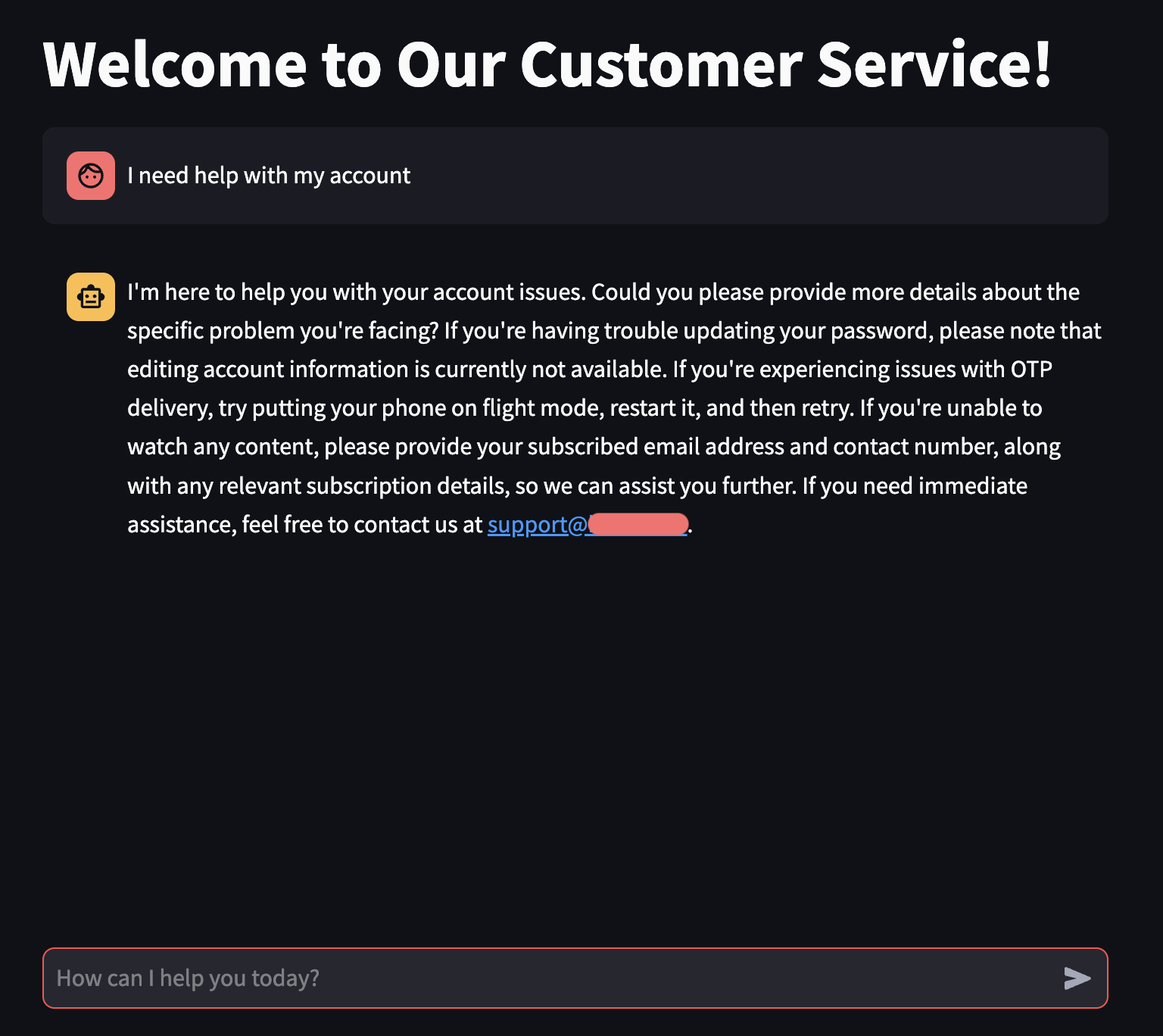}
    \small{(c)}
  \end{minipage}
  \caption{Screenshots of the chatbot implementation in Streamlit: (a) Home screen of the chatbot, showing three FAQs; (b) Answer printed from the FAQ database; (c) Answer generated through \textit{GPT-4o} based on the user question and the context retrieved from the FAQ database}\label{fig4}
  \Description{Screenshots of the chatbot implementation in Streamlit: (a) Home screen of the chatbot, showing three FAQs; (b) Answer printed from the FAQ database; (c) Answer generated through \textit{GPT-4o} based on the user question and the context retrieved from the FAQ database.}
\end{figure*}

Figure~\ref{fig4} shows three screenshots of BanglAssist's interface implementation in Streamlit: (a) the home screen of the chatbot, showing three FAQs the user can choose from; (b) a chatbot answer to an FAQ, printed from the FAQ database; (c) a chatbot answer generated through \textit{GPT-4o} based on the user question and the context retrieved from the FAQ database.

\clearpage
\newpage

\twocolumn

\section{BanglAssist Evaluation Queries Breakdown}
\label{appendix:e}

\begin{table}[H]
    \caption{Language and topic of each query used to evaluate BanglAssist}
    \label{tab3}
    \begin{tabular}{lcl}
        \toprule
        \textbf{Query} & \textbf{Language} & \textbf{Topic}\\
        \toprule
        1 & Bengali & Unable to watch specific episode\\
        2 & Bengali & Discount not applied to total amount\\
        3 & Bengali & Request to add specific content\\
        4 & Bengali & Poor video quality while streaming\\
        5 & Bengali & Account idle despite subscription\\
        6 & Banglish & Content streamed on illegal websites\\
        7 & Bengali & Black screen while playing video\\
        8 & English & Issue downloading specific content\\
        9 & Banglish & Too many devices connected to account\\
        10 & English & Cancel subscription via Google Play\\
        11 & Banglish & Problem receiving OTP to login\\
        12 & Banglish & Auto-pay showing wrong renewal date\\
        13 & English & Problem receiving OTP on new phone\\
        14 & Banglish & Wrong period showing for offer\\
        15 & English & Dissatisfaction with released content\\
        16 & Banglish & Problem accessing platform\\
        17 & Banglish & Issue while streaming content\\
        18 & Banglish & Issue while streaming specific content\\
        19 & English & No video while streaming content\\
        20 & Banglish & Content leaked via Telegram\\
        \bottomrule
    \end{tabular}
\end{table}

\section{Equations}
\label{appendix:f}

In the following, \textit{|Q|} is the total number of queries, MRR indicates the Mean Reciprocal Rank, and $Rank_{i}$ is the position of the first relevant document for query \textit{i}.

\begin{equation} \label{eq1}
    \text{Precision@k} = \frac{1}{|Q|} \sum_{i=1}^{|Q|} \frac{\text{Number of relevant docs in top } k}{k}
\end{equation}

\begin{equation} \label{eq2}
    \text{Recall@k} = \frac{1}{|Q|} \sum_{i=1}^{|Q|} \frac{\text{Number of relevant docs in top } k}{\text{Total number of relevant docs}}
\end{equation}

\begin{equation} \label{eq3}
    \text{MRR@k} = \frac{1}{|Q|} \sum_{i=1}^{|Q|} \frac{1}{\text{Rank}_i}, \quad \text{where } \text{Rank}_i \leq k 
\end{equation}

\section{BanglAssist Evaluation Breakdown by Language}
\label{appendix:g}

\begin{table}[H]
    \caption{Average retrieval scores per language before reranking (\textit{k = 5} and \textit{k = 3}) and after reranking (\textit{k = 3})}
    \label{tab4}
    \begin{tabular}{lccc}
        \toprule
        \textbf{Language} & \textbf{Precision@k} & \textbf{Recall@k} & \textbf{MRR@k}\\
        \toprule
        \multicolumn{4}{l}{\textit{k = 5: Retrieval performance evaluation}}\\
        \multicolumn{4}{l}{\textit{\textbf{(Before reranking)}}}\\
        \midrule
        Bengali & \textbf{0.67} & \textbf{0.50} & \textbf{1.00}\\
        English & 0.64 & 0.39 & 0.80\\
        Banglish & 0.47 & 0.40 & 0.81\\
        \midrule
        \multicolumn{4}{l}{\textit{k = 3: Reranking performance evaluation}}\\
        \multicolumn{4}{l}{\textit{\textbf{(Before and after reranking)}}}\\
        \midrule
        Bengali & \textbf{0.89} & \textbf{0.43} & \textbf{1.00}\\
        English & 0.73 & 0.30 & 0.80\\
        Banglish & 0.67 & 0.35 & 0.81\\
        \midrule
        Bengali & \textbf{0.78} & \textbf{0.42} & \textbf{1.00}\\
        English & 0.73 & 0.30 & 0.80\\
        Banglish & 0.61 & 0.35 & 0.83\\
        \bottomrule
    \end{tabular}
\end{table}
\end{document}